\documentclass[conference]{IEEEtran}
\IEEEoverridecommandlockouts
\usepackage{cite}
\usepackage{amsmath,amssymb,amsfonts}
\usepackage{algorithmic}
\usepackage{graphicx}
\usepackage{epstopdf}
\usepackage{textcomp}
\usepackage{xcolor}
\def\BibTeX{{\rm B\kern-.05em{\sc i\kern-.025em b}\kern-.08em
    T\kern-.1667em\lower.7ex\hbox{E}\kern-.125emX}}
\begin{document}

\title{Robust Vector Perturbation Precoding Design for MIMO Broadcast Channel    \\
\thanks{This work was supported by State Key Laboratory of Wireless Mobile Communications, China Academy of Telecommunication Technology(CATT).}
}

\author{\IEEEauthorblockN{Liutong Du, Lihua Li, Ping Zhang}
\IEEEauthorblockA{\textit{State Key Laboratory of Networking and Switching Technology} \\
\textit{Beijing University of Posts and Telecommunications}\\
Beijing 100876, P.R. China \\
Email:\{liutongdu,lilihua\}@bupt.edu.cn}
}

\maketitle

\begin{abstract}
We consider the vector perturbation (VP) precoder design for multiuser multiple-input single output (MU-MISO) broadcast channel systems which is robust to power scaling factor errors. VP precoding has so far been developed and analyzed under the assumption that receivers could have known the power scaling factor in advance of tranmission perfectly, which is hard to obtain due to the large dynamic range and limited feedforward. However, as demonstrated in our results the performance of VP precoding is quite sensitive to the accuracy of power scaling factor and always encounter an error floor at mid to high signal-to-noise ratio (SNR) regimes. Motivated by such observations, we propose a robust VP precoder based on the minimum  mean square error (MMSE) criterion. Simulation results show that, the robust VP precoder outperforms conventional VP precoding designs, as less sensitive to power scaling factor errors.    
\end{abstract}


\section{Introduction}
It is well known that multiple-input multiple output (MIMO) systems can provide higher sum rates compared with single antenna systems \cite{b1}. For the downlink of MIMO, as it can serve more than one user simultaneously, is so-called broadcast channel (BC). In paper \cite{b2}, dirty paper coding (DPC) \cite{b3} is proved optimal to access the sum capacity of MIMO BC, but there are many challenges to put DPC into practice, to circumvent the problem, many suboptimal precoding techniques has been proposed. These techniques mainly lies into two categories: linear precoding and non-linear precoding. Zero-forcing (ZF) \cite{b4} precoding (or channel inversion) and minimum mean square error (MMSE) \cite{b5} are among the most popular linear precoding schemes, which has lower complexity but suffer from a performance loss compared with non-linear precoding schemes \cite{b6}. Vector perturbation (VP) precoding \cite{b7} and Tomlinson-Harashima (TH) precoding \cite{b8} are two representative non-linear precoding techniques. In \cite{b9}, it has been proven that VP precoding can achieve full diversity. 

In conventional VP precoding (CVP)\cite{b7}, the transmitter adds a perturbation vector to the modulated data vector and generates the transmit vector by multiplying the perturbed vector with a precoding matrix. The transmitter selects the precoding matrix to mitigate the inter-user interference and solve the perturbation vector under the criterion of reducing the unscaled transmit power with so-called sphere decoder \cite{b7},\cite{b10}. The receiver recovers the data vector indiviually by multiplying a power scaling factor and then pass the scaled vector through a modulo operator to eliminate the effect of perturbed vector.  In \cite{b11}, an improved VP precoding scheme namely MMSE-VP was proposed to jointly optimize the precoding matrix and perturbation vector under the criterion of minimizing the mean square error between the perturbed data vector and received scaled vector. 

 In \cite{b12,b13}, robust VP precoding schemes are designed to take imperfections on transmit channel  state information (CSI) into consideration, such as quantized channel feedback\cite{b12}, or separately regards CSI as quantized channel direction information (CDI) and channel magnitude information (CMI) \cite{b13}, it turns out that CMI performs an rather important role in non-linear precoding VP compared to linear precoding ZF. As far as we know, the existing robust VP precoders always assume that the receivers can perfectly know the power scaling factor in advance, which is hard to obtain in practice. In \cite{b14}, a VP precoder was proposed as the receivers do not require the power scaling factor, but it only works with a specific modulation scheme, which is not appliable for modern communication. In \cite{bb}, the proposed VP precoder would discard the data vector which would cause a large dynamic of power scaling factors and thus keep the power scaling factors with a fixed range that the receivers have known in advance, however it is too ideal and the performance would degrades badly if the outrage takes place frequently. Hard and important as it is to obtain the correct power scaling factors, which is data- and channel- dependent\cite{b7}, it is reasonable to deliver it through a limited feedforward link. As the time and frequency resource is limited, we can not deliver all power factors to the receiver, the delivered power factor also have to suffer from limited quantization, together with time delay and other impfectness.  Motivated by these observations, we propose a robust VP precoder which is less sensitive to the accuracy of power scaling factors at the receivers within MMSE criterion. Simulation results show that with a fixed power scaling error, as the performance of  MMSE-VP\cite{b11} converges to CVP \cite{b7} eventually with increased SNR, the proposed scheme outperforms CVP with a lower error floor at high SNR regimes and shares almost the same performance with MMSE-VP at mid-to-high SNR, and is diversity gain\cite{book} achievable with a varied noise-adaptive power scaling error.

The rest of this paper is organized as follows. Section ~\ref{sys} introduces the MIMO system model and the procedure of conventional VP precoding, furthermore we introduced the MMSE-based VP precoder. In section III, we give the error model of power scaling factor, and propose the robust VP precoding scheme based on given error model within MMSE criterion. In section IV, simulation results are presented. Section V concludes this paper.

\textbf{Notations:}
In this paper, super scripts ${\left( {\bf{A}} \right)^T}$, ${\left( {\bf{A}} \right)^H}$ denote the transpose, the conjugate transpose of matrix ${\bf{A}}$. $\left\| {\bf{x}} \right\|$ denotes the Euclidean norm of vector ${\bf{x}}$. ${{\bf{I}}_M}$ is the $M \times M$ identity matrix. ${E_n}$ stands for taking expectation over $n$. $\Re \left( c \right)$, $\Im \left( c \right)$ denotes the real and imaginary part of $c$. $\left\lfloor  \bullet  \right\rfloor $ denotes the floor operation.

\begin{figure}[htbp]
\includegraphics[width = 3.5in]{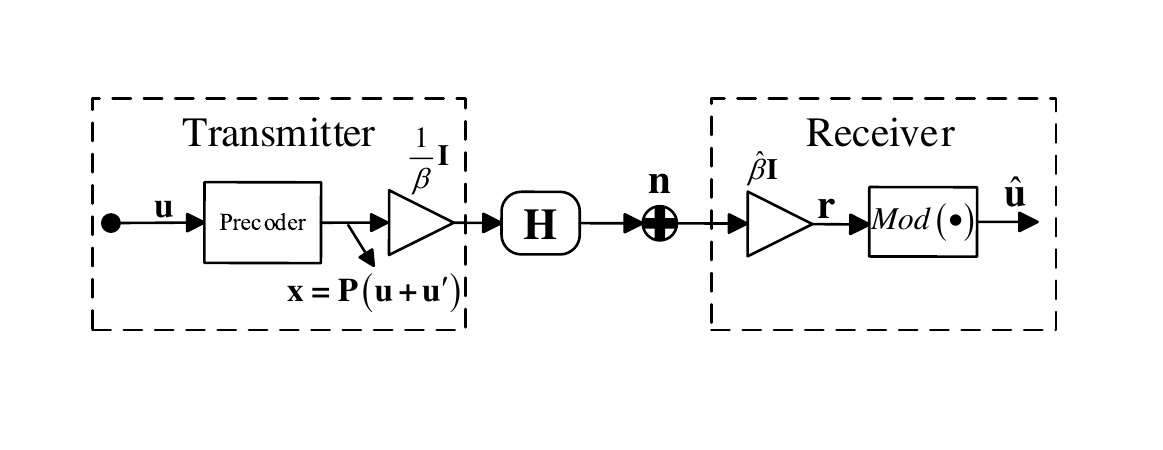}
\caption{MU-MISO system with VP precoding}
\label{fig1}
\end{figure}
\section{SYSTEM OVERVIEW}
\label{sys}
\subsection{MIMO system overview}
We assume the base station deploys $N_t$ antennas and serves $N_r$ non-cooperative single antenna users simultaneously, which follows  ${N_r} \le {N_t}$ . Channel vector of user    $i$ is denoted as ${\bf{h}}_i = {\left[ {{h_{i1}}, \cdots ,{h_{i{N_t}}}} \right]^T}$, where   ${h_{i,j}}$  signifies the channel gain between the  $j$th transmit antenna and   $i$th user. In this paper, we suppose the channel gain to be independent and identically distributed (i.i.d) complex Gaussian random variables with zero mean and unitary variance. The receive signal at user $i$ is given by
\begin{equation}
{{{y}}_i} = {\bf{h}}_i^T{\bf{x}} + {{{n}}_i}\label{eq1}
\end{equation}

Where  ${\bf{x}}$ is the transmit vector,  ${n_i}$ is the zero-mean complex Gaussian noise at user $i$ , which has a variance of   $\sigma _n^2$ . Under the unit transmit power constraint, we can define SNR as
\begin{equation}
 \rho  = \frac{1}{{\sigma _n^2}}\label{eq2}   
\end{equation}

By stacking the received signal of ${N_r}$
 users, we can obtain the received vector  as
\begin{equation}
 {\bf{y = Hx + n}}\label{eq3}   
\end{equation}
Where ${\bf{H}}$ is the channel matrix, ${\bf{n}}$ is the additive noise vector.
At the base station, bit data stream ${\bf{a}}$ generates the symbol data vector ${\bf{u}} = {\left[ {{u_1}, \cdots ,{u_{{N_r}}}} \right]^T}$ by quadrature amplitude modulation (QAM). In \cite{b6}, Christian et.al point out that linear precoding techniques suffer a severe performance loss under full-load scrnario ($N_r = N_t$) and can not access the sum capacity which  grows linearly with the minimum of the number of base station antennas and users, and they introduced the VP precoding scheme in \cite{b7}, which could achieve a near capacity sum rates of MU-MIMO BC.\\
\subsection{Conventional VP precoding}
The system diagram of VP precoding is shown in Fig.~\ref{fig1}. Note that, there are more than one receivers in MU-MISO systems which can perform signal detection independently, here we draw only one receiver just to demonstrate the entire tansmit and receive procedure. The basic idea of VP is to reduce the transmit power by perturb the data vector ${\bf{u}}$ with a scaled Gaussian integers ${\bf{u'}}$, which is so called \textit{perturbation vector}. The perturbation vector ${\bf{u'}}$ is selected to minimize the unscaled power of perturbed data vector
\begin{equation}
	{\mathbf{u'}} = \mathop {\arg \min }\limits_{{\mathbf{\hat u}}} {\left\| {{\mathbf{P}}\left( {{\mathbf{u + \hat u}}} \right)} \right\|^2}\label{eq4}
\end{equation}

The scale factor $\tau$ is determined by the modulation scheme, throughout this paper, we follow the set in \cite{b7}, in which $\tau$ is set as $\tau  = 2\left( {{c_{\max }} + {\Delta  \mathord{\left/
 {\vphantom {\Delta  2}} \right.
 \kern-\nulldelimiterspace} 2}} \right)$, where ${c_{\max }}$ is the absolute value of the constellation symbol(s) with largest magnitude, and $\Delta$ is the spacing between constellation points. Then the transmit vector ${\bf{x}}$ is formed by multiplying the perturbed data vector $\left( {{\bf{u + u'}}} \right)$ with a ${N_t} \times {N_r}$ precoding matrix ${\bf{P}}$, and is normalized to unit power at base station to follow the transmit power limit as
 \begin{equation}
  {\bf{x}} = \frac{1}{\beta }{\bf{P}}\left( {{\bf{u}} + {\bf{u'}}} \right)  \label{eq5}
 \end{equation} 
 
Where $\beta  = \sqrt {{{\left\| {{\bf{P}}\left( {{\bf{u}} + {\bf{u'}}} \right)} \right\|}^2}} $ is the power scaling factor. It is obvious that $\beta$ is determined not only by precoding matrix ${\bf{P}}$ but also by data vector ${\bf{u}}$.The perturbation vector ${\bf{u'}}$, which is also detemined by the precoding matrix ${\bf{P}}$ and data vector ${\bf{u}}$, can be solved via so-called \textit{sphere-decoder}. Substituting \eqref{eq5} into \eqref{eq3}, the received vector can be obtained as 
\begin{equation}
  {\bf{y}} = \frac{1}{\beta }{\bf{HP}}\left( {{\bf{u}} + {\bf{u'}}} \right) + {\bf{n}}  \label{eq6}
\end{equation}
It is assumed that all receivers could have known $\beta$ perfectly in advance of transmission in \cite{b7}, the transmitter is assumed to have prefect knowledge of channel information in advance of transmission as well. The received signal vector recovered with $\beta$ is denoted as
\begin{equation}
\begin{aligned}
{\mathbf{r}} &= \beta {\mathbf{y}} \\ 
&= {\mathbf{HP}}\left( {{\mathbf{u + u'}}} \right) + \beta {\mathbf{n}} \\ 
\end{aligned}\label{eqy}
\end{equation}
 User $i$ estimates its data symbol by performing a modulo operation on the scaled received symbol as
\begin{equation}
    {\hat u_i} = M\left( r_i \right)\label{eq7}
\end{equation}
The complex modulo operator, which translates $r_i$ into the aimed constellation region can be defined as follows:
\begin{equation}
    M(a) = a - \left\lfloor {\frac{{\Re (a)}}{\tau } + \frac{1}{2}} \right\rfloor \tau  - j\left\lfloor {\frac{{\Im (a)}}{\tau } + \frac{1}{2}} \right\rfloor \tau  \in \Lambda \label{eq8}
\end{equation}
Where the constellation region $\Lambda$ is defined as follows:
\begin{equation}
 {\Lambda  = \left\{ {c \in \mathbb{C}| - \frac{\tau }{2} \leqslant \Re \left( c \right) < \frac{\tau }{2},- \frac{\tau}{2} \leqslant \Im \left( c \right) < \frac{\tau }{2}} \right\}}.\label{eq9}
 \end{equation}
\begin{figure}[htbp!]
 	\includegraphics[width = 3in]{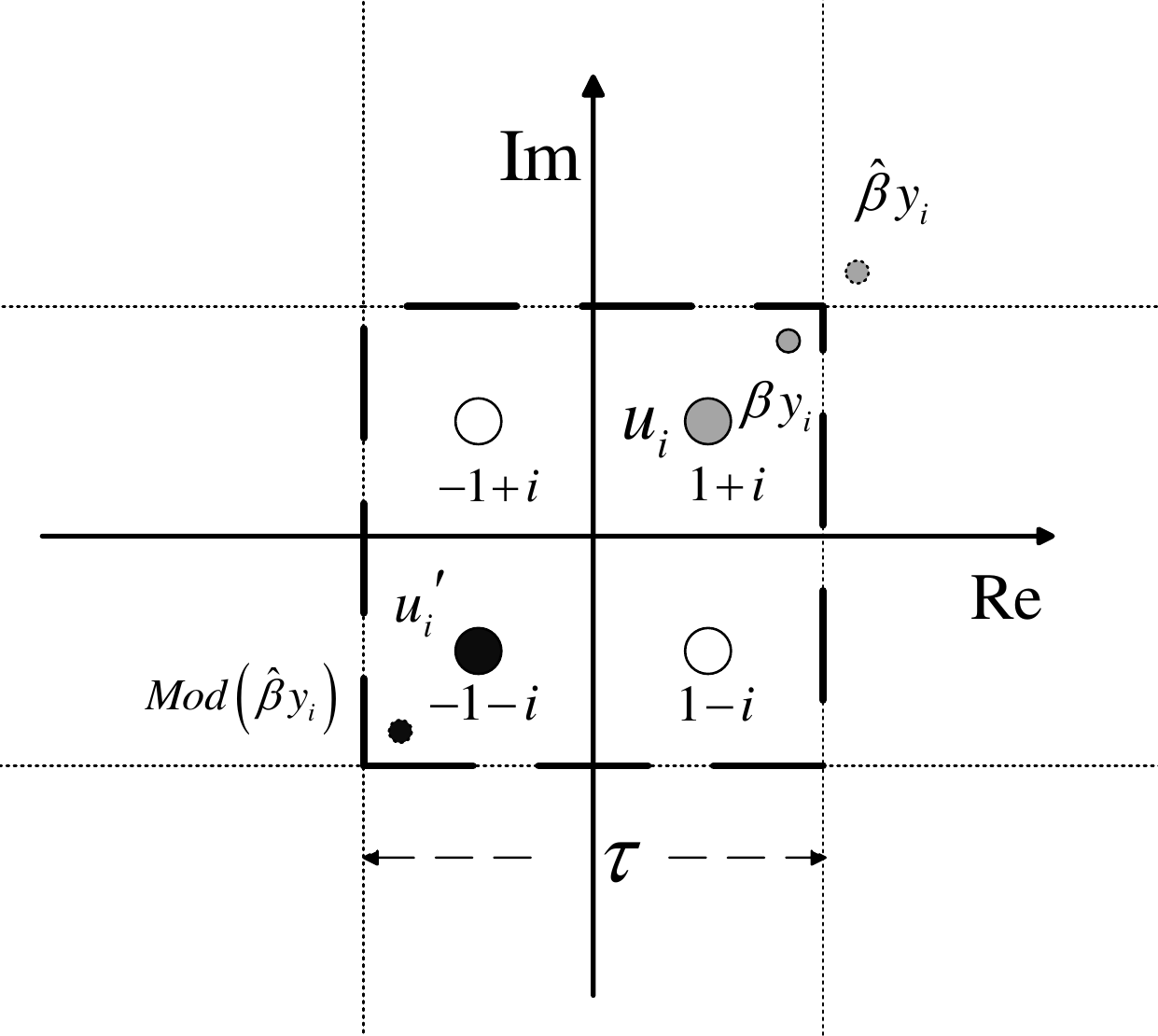}
 	\caption{Modulo process in VP precoding with 4QAM}
 	\label{mod}
\end{figure} 

 After modulo process, both the real part and the imaginary part of the operand were mapped to the interval $\left[ { - \frac{\tau }{2},\frac{\tau }{2}} \right)$, and the demodulator selects the constellation with nearest Euclidean distance and maps it to the correspondence bit sequence.  Fig.~\ref{mod} shows the process of modulo operator and how would power scaling factor affect the demodulation result. We represent the modulated data symbol of user $i$ as $u_i$, if the receiver scales the symbol correctly with the right factor $\beta$, after demodulation of noise-distorted symbol $\beta{y_i}$, the user would recover the symbol as '$1+i$'. But if the receiver could only get an inaccurate power scaling factor $\hat \beta$, which may be really close to $\beta$, then the received symbol after modulo process would be mapped to ${u_i}^\prime $, which locates at the decision region of '$-1-i$'. We can see from this example that unlike noise which always distort the modulated symbol to the neighbouring symbol, with an incorrect power scaling factor, severer mistakes would take place. \\
 \subsection{MMSE VP precoding}
To introduce the MMSE criterion based VP, which is also called as \textit{Wiener filter}\cite{wf} (WF) VP, we first define a deviation vector, which measures the distortion between the scaled received vector $\beta {\bf{y}}$ and the perturbed data vector $\left( {{\bf{u + u'}}} \right)$, as
\begin{equation}
    {\bf{d}} = \beta {\bf{y}} - \left( {{\bf{u + u'}}} \right) = \left( {{\bf{HP - }}{{\bf{I}}_{{N_r}}}} \right)\left( {{\bf{u + u'}}} \right) + \beta {\bf{n}}\label{eq10}
\end{equation}
Given the data vector ${\bf{u}}$ and channel matrix ${\bf{H}}$, the mean square error (MSE) as a function of ${\bf{u'}}$ and ${\bf{P}}$ is expressed as 
\begin{equation}
    \begin{array}{c}
\begin{gathered}
  e\left( {{\mathbf{u'}},{\mathbf{P}}} \right) \hfill \\
   = {E_{\mathbf{n}}}\left( {{{\left\| d \right\|}^2}|{\mathbf{H}},{\mathbf{u}}} \right) \hfill \\
   = {\left\| {\left( {{\mathbf{HP}} - {{\mathbf{I}}_{{N_r}}}} \right)\left( {{\mathbf{u}} + {\mathbf{u'}}} \right)} \right\|^2} + {N_r}{\beta ^2}\sigma _n^2 \hfill \\ 
\end{gathered} \label{eq11}
\end{array}
\end{equation}
The optimal precoding matrix ${{\bf{P}}_{o}}$ that minimizes \eqref{eq11} is obtained following \cite{b11}. First, we assume the optimum perturbation vector ${{{\mathbf{u'}}}_o}$ is given and optimise over ${\mathbf{x}}$ and $\beta$ using the Lagrangian approach, which will lead to a unique solution as the global optimum for fixed ${{{\mathbf{u'}}}_o}$. Then, we further minimise the MSE by searching over ${{{\mathbf{u'}}}_o}$ under the assumption that the optimum  ${\mathbf{x}}$ and $\beta$ for the respective ${{{\mathbf{u'}}}_o}$ is employed. This procedure would lead to the following optimum solution:
\begin{equation}
    {{\bf{P}}_{o}} = {{\bf{H}}^H}{\left( {{\bf{H}}{{\bf{H}}^H} + {N_r}\sigma _n^2{{\bf{I}}_{{N_r}}}} \right)^{ - 1}}\label{eq12}
\end{equation}
The optimal perturbation vector ${\bf{u'}}$ can be obtained as 
\begin{equation}
{{{\mathbf{u'}}}_o} = \mathop {\arg \min }\limits_{u'} {N_r}\sigma _n^2{{\mathbf{s}}^H}{\left( {{\mathbf{H}}{{\mathbf{H}}^H} + {N_r}\sigma _n^2{{\mathbf{I}}_{{N_r}}}} \right)^{ - 1}}{\mathbf{s}}
\label{eq13}
\end{equation}
Here we use ${\mathbf{s}} = \left( {{\mathbf{u}} + {\mathbf{u'}}} \right)$ to denote the perturbed signal vector for short. With Cholesky factorozation, \eqref{eq13} can be rewritten as
\begin{equation}
{{\mathbf{u'}}_o} = \mathop {\arg \min }\limits_{u'} {\left\| {{\mathbf{L}}\left( {{\mathbf{u + u'}}} \right)} \right\|^2}
\label{eqL}
\end{equation}
Where ${\left( {{\mathbf{H}}{{\mathbf{H}}^H} + {N_r}\sigma _n^2{{\mathbf{I}}_{{N_r}}}} \right)^{ - 1}} = {{\mathbf{L}}^H}{\mathbf{L}}$, ${\mathbf{L}}$ is an up-trianglar matrix.Compare\eqref{eqL} with \eqref{eq4}, both perturbation vectors $\mathbf{u'}$ and ${\mathbf{u'}}_o$ can be solved with sphere decoder, as they shared the same form.

\section{ROBUST VECTOR PERTURBATION PRECODING DESIGN}
In order to focus on the robust design of power factor imperfection, we assume transmitter can obtain the channel information ideally through reciprocal channels in this paper, e.g., we always assume the up-link and down-link channel are identical in time division duplex (TDD) systems\cite{b15}, where the base station can get perfect channel state information through up-link training.From Section ~\ref{sys}, we can learn that optimal precoding matrix ${{\bf{P}}_{o}}$ is determined only by the channel matrix $\bf{H}$ given the fixed ${N_r}$ and $\sigma _n^2$. Together with \eqref{eq5}, \eqref{eq12}, \eqref{eq13}, we can further figure out the power scaling factor $\beta$ is both channel- and data- dependent. Nevertheless, from \cite{bb}, we can learn that $\beta$ have a wide dynamic range. As in standards like 4G long-term evolution (LTE), the base station could only send limited number of signaling symbols like cell-specific reference signal (CRS) and demodulation reference signal (DMRS) through a so called \textit{feedforward} link. The power scaling factors obtained at transmitter can only be sent via a low rate feedforward control channel. We model the power scaling factor error within a limited scenario for two  reasons: first, it is cost expensive to deliver power scaling factors for every transmit data vector, it is unavoidable to transmit the average value (or other forms) of power scaling factor for the entire resource block (RB)(or other sizes) to reduce the total overhead; second, the transmitted power factor is represented by limited number of bits, which introduces the quantization error. 
Then, we represent the quantized power scaling factor that user received as $\hat \beta$, and the error between $\beta $ and $\hat\beta$ can be expressed as:
\begin{equation}
    {\Delta \beta  = \beta  - \hat \beta} \label{eq14}
\end{equation}
In this paper, we assume that the relative power factor error $\frac{{\Delta \beta }}{\beta } = \frac{{\beta  - \hat \beta }}{\beta }$ is a Gaussian random variable that follows independent zero-mean Gaussian distribution with a variance of $\sigma _q^2$. To better focus on the imperfections of $\beta$ we assume the transmitter could get perfect CSI through channel reciprocity of TDD system, and the receivers get the imperfect power scaling scalar $\hat \beta $ via an error-free and delay-free link.
Then the received signal vector can expressed as:
\begin{equation}
    {\bf{y}} = \frac{1}{\beta }{\bf{HP}}\left( {{\bf{u}} + {\bf{u'}}} \right) + {\bf{n}}\label{eq15}
\end{equation}
As the receivers can only get an incorrect power scaling factor $\hat \beta$, the received signal vector  recovered with $\hat \beta$ can be denoted as
\begin{equation}
    {\bf{r}} = \hat \beta {\bf{y}} = \frac{{\hat \beta }}{\beta }{\bf{HP}}\left( {{\bf{u}} + {\bf{u'}}} \right) + \hat \beta {\bf{n}}\label{eq16}
\end{equation}
The deviation vector of recovered signal $\bf{r}$ and perturbed transmit vector $\bf{u}+\bf{u'}$ can be obtained as 
\begin{equation}
  \begin{array}{l}
  {\mathbf{d}} = {\mathbf{r}} - \left( {{\mathbf{u}} + {\mathbf{u'}}} \right) \\ 
  = \left( {{\mathbf{HP}} - {\mathbf{I}}} \right)\left( {{\mathbf{u}} + {\mathbf{u'}}} \right) - \frac{{\Delta \beta }}{\beta }{\mathbf{HP}}\left( {{\mathbf{u}} + {\mathbf{u'}}} \right) + \frac{{\hat \beta }}{\beta }\beta {\mathbf{n}} \\ 
  \end{array} \label{eq17}
\end{equation}
As precoding is done at transmitter, given data vector ${\bf{u}}$ and channel information ${\bf{H}}$, the MSE as a function of perturbation vector ${\bf{u'}}$ and precoding matrix ${\bf{P}}$ is obtained by taking expectation over noise ${\bf{n}}$ and the power factor error $\Delta \beta $ as
\begin{equation}
    \begin{array}{l}
e\left( {P,u'} \right)\\
 = {E_{{\bf{n}},\Delta \beta }}\left( {{{\left\| {\bf{d}} \right\|}^2}|{\bf{u}},{\bf{H}},\hat \beta } \right)\\\label{eq18}
 = {\left\| {\left( {{\bf{HP}} - {\bf{I}}} \right)\left( {{\bf{u}} + {\bf{u'}}} \right)} \right\|^2} + {N_r}{\beta ^2}\left( {\sigma _q^2 + \sigma _n^2(1 + \sigma _q^2)} \right)
\end{array}
\end{equation}
Together with \eqref{eq10} and \eqref{eq18}, it is clear that the optimization problem shares the same form with MMSE-VP precoding. As fully studied in paper \cite{b11} and \cite{b13}, we can further conclude the robust precoding matrix ${{\bf{P}}_q}$ without perfect power factor at receivers as
\begin{equation}
{{\bf{P}}_q}{\bf{ = }}{{\bf{H}}^{{H}}}{\left( {{\bf{H}}{{\bf{H}}^{{H}}} + {N_r}\left( {\sigma _q^2 + \sigma _n^2(1 + \sigma _q^2)} \right){{\bf{I}}_{N_r}}} \right)^{ - 1}}\label{eq19}
\end{equation}
Then, with precoding matrix, we can solve the optimal perturbation vector with MMSE criterion as
\begin{equation}
   \begin{gathered}
  {{{\mathbf{u'}}}_q} = \mathop {\arg \min }\limits_{{\mathbf{\hat u}}} {N_r}\left( {\sigma _q^2 + \sigma _n^2(1 + \sigma _q^2)} \right){\left( {{\mathbf{u + u'}}} \right)^H} \\ 
   \times {\left( {{\mathbf{H}}{{\mathbf{H}}^H} + {N_r}\left( {\sigma _q^2 + \sigma _n^2(1 + \sigma _q^2)} \right){{\mathbf{I}}_{{N_r}}}} \right)^{ - 1}}\left( {{\mathbf{u + u'}}} \right) \\ 
\end{gathered} \label{eq20}
\end{equation}
Furthermore, with Cholesky factorization, we can get an up-trianglar matrix ${\bf{L}}$ which fulfills
\begin{equation}
    {\left( {{\mathbf{H}}{{\mathbf{H}}^H} + {N_r}\left( {\sigma _q^2 + \sigma _n^2(1 + \sigma _q^2)} \right){{\mathbf{I}}_{{N_r}}}} \right)^{ - 1}} = {{\bf{L}}^H}{\bf{L}}\label{eq21}
\end{equation}
Then, with \eqref{eq20} and \eqref{eq21}, we can further get the perturbation vector ${{\bf{u'}}_q}$ as in \cite{b11} 
\begin{equation}
    {{\bf{u'}}_q} = \mathop {\arg \min }\limits_{{\bf{\hat u}}} \left\| {{\bf{L}}\left( {{\bf{u + \hat u}}} \right)} \right\|_2^2 \label{eq22}
\end{equation}
Compare \eqref{eq22} with \eqref{eq4}, we can see that, the problem of solving perturbation vector for robust VP precoding and conventional VP shares the same form. This problem can be solved with sphere decoder as proposed in \cite{b7}, in this paper we use the quick sphere decoding algorithm introduced by \cite{b10}, which avoid redundant calculation, and achieve the exact performance at the same time.

\begin{figure}[htbp]
\includegraphics[width = 3.2in]{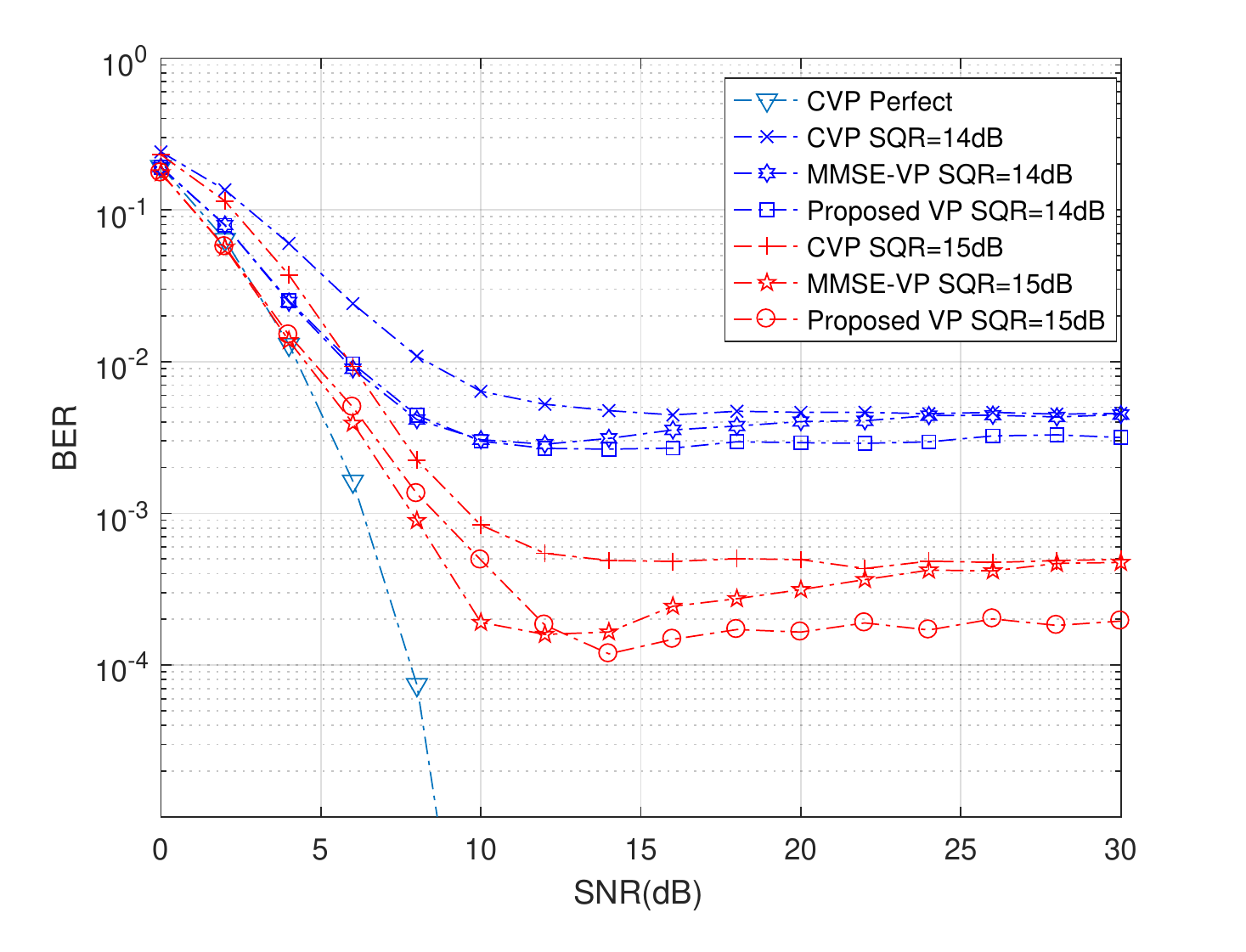}
\caption{BER performance of different  VP precoding schemes with the fixed power scaling factor error}
\label{BER}
\end{figure}
\section{SIMULATION RESULTS}
In this section, we evaluate the performance of the proposed robust VP precoding scheme with conventional VP\cite{b7} and MMSE-VP\cite{b11}. We consider a MU-MISO BC channel with a base station equipped with $N_t=4$ transmit antennas serving $N_r=2$ single antenna users at the same time. We note that, the proposed scheme can be applied to any MU-MISO system with ${N_r} \le {N_t}$. To better demonstrate the affections of power scaling factor error, we define the signal-to-quantization error ratio (SQR) as $SQR(dB) = 10lo{g_{10}}\left( {{1 \mathord{\left/
			{\vphantom {1 {\sigma _q^2}}} \right.
			\kern-\nulldelimiterspace} {\sigma _q^2}}} \right)$.
   
Performance is evaluated in terms of coded bit error rate (BER) versus signal-to-noise ratio (SNR) with ${1 \mathord{\left/
 {\vphantom {1 2}} \right.
 \kern-\nulldelimiterspace} 2}$ rate Turbo code. The modulation scheme is 16QAM. We average the performance over $10000$ transmission time intervals (TTI) within 3GPP channel models\cite{b16}. The environment is set as 'urban macro-cell'. \\
\indent In Fig.~\ref{BER}, conventional VP, MMSE-VP, and proposed VP under different SQRs are compared with conventional VP with no power scaling factor error. The SQR is set as 14 dB and 15 dB as in our former simlulations, it turns out that, as we deliver 6 or 12 power factors per RB, the statistical average SQRs are aroud 14-15dB. It can be seen in Fig.~\ref{BER} that the proposed robust VP outperforms conventional VP over all SNR regimes as it considers both noise and power scaling factor error. We can further conclude from Fig.~\ref{BER} that the performance of VP relies heavily on the accuracy of power scaling factors as they all encounter an error floor at high SNR with a fixed SQR. It is clear that all these precoding schemes are interference-limited at high SNR region, but the proposed robust VP precoder can provide a better performance.

\begin{figure}[!h]
	\includegraphics[width = 3.42in]{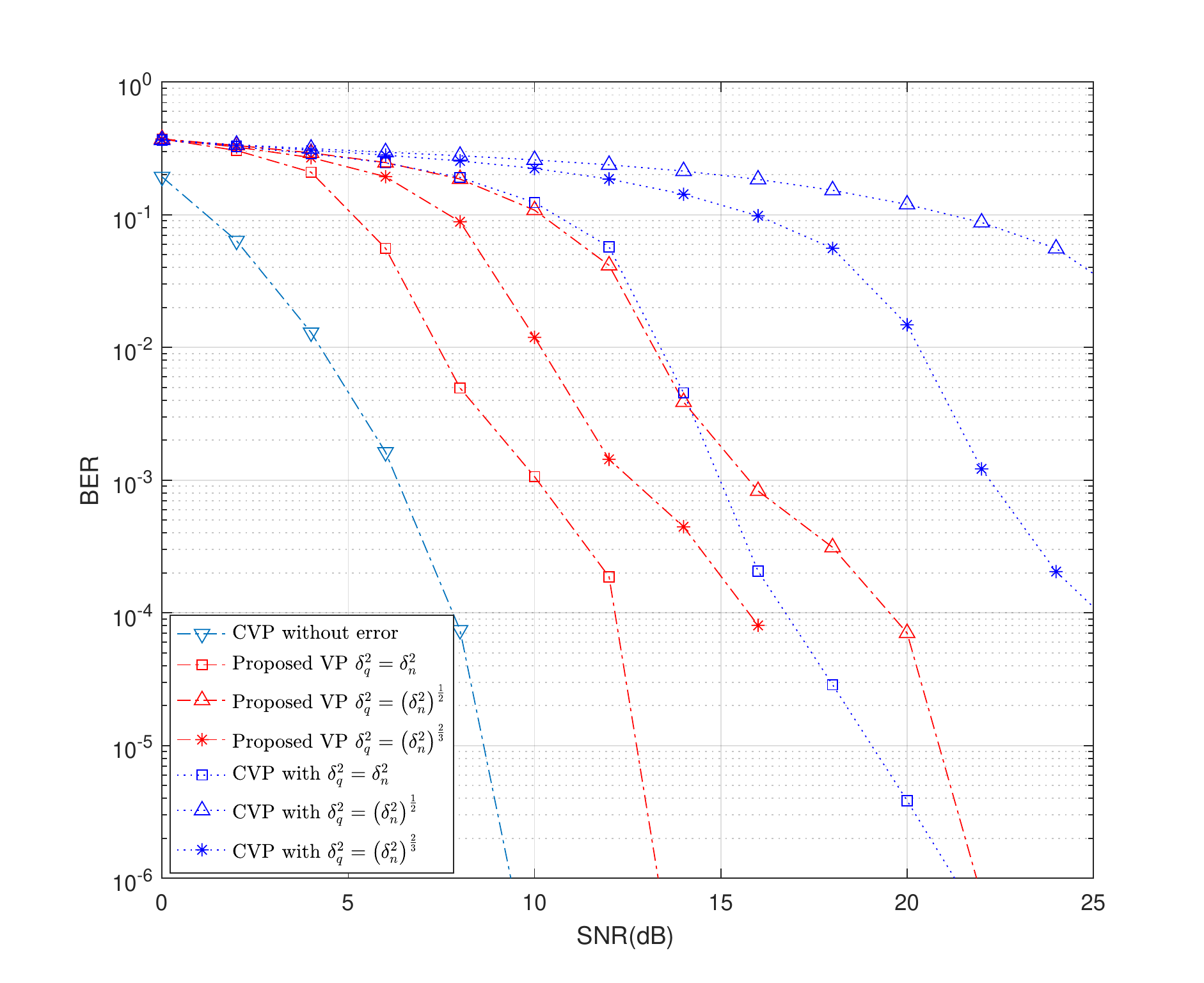}
	\caption{BER performance of VP precoding with power scaling factor error varied with SNR}
	\label{BER2}
\end{figure}

\indent As shown in Fig.~\ref{BER}, the performance of VP is limited with a fixed SQR. It is interesting to consider whether the diversity gain could be achieved with power scaling factor errors. As we always presume a better CSI could be obtainded with a higher SNR \cite{b12}, similarly, we further assume the relative variance $\sigma_q^2$ as a function of SNR, 
 for example, $\sigma_q^2 = \sigma_n^2$. 
Note here, with ${\text{0}} < \sigma _n^2 \leqslant 1$, we always have $\sigma _n^2 \leqslant {\left( {\sigma _n^2} \right)^{\frac{2}{3}}} \leqslant {\left( {\sigma _n^2} \right)^{\frac{1}{2}}}$. 
 Under such assumption, the BER performance of robust VP  and CVP is shown in Fig.~\ref{BER2}. The proposed VP scheme is more robust than CVP with a varied power scaling factor error. With $\sigma_q^2 = \sigma_n^2$, the performance gap between robust VP and CVP without error is about 3dB at $10^{-3}$. Both Fig.~\ref{BER} and Fig.~\ref{BER2} give valuable guidelines for system design. For instance, we can further figure out how many power scaling factors are needed per RB if we want to keep SQR under a certain degree to achieve a better BER performance or how many bits it will cost to reap the diversity gain.
 
 As we can see from \eqref{eq12} and \eqref{eq19}, the proposed VP and MMSE-VP shares the same form of precoding matrix thus have the same complexity $\mathcal{O}({N_r}^3)$ as CVP. As the computation complexity of serial Cholesky factorization is $\mathcal{O}({N_r}^3)$ and $\mathcal{O}({N_r})$ for parallel Cholesky factorization\cite{b17}, thus these schemes will have  the same computation complexity within a selected sphere decoder.
\section{CONCLUSION} 
In this paper, we study the robust VP precoding scheme design for which the receivers can not get power scaling factors ideally. Based on the derived closed form of the mean square error, we consider a joint optimal design of precoding matrix and perturbation vector under MMSE criterion. The simulation results have shown that performance of VP precoding depends critically on the accuracy of power scaling factors as they all encounter an error floor which is caused by the interference introduced by the imperfections of power scaling factor delivery. The proposed robust VP precoder can always achieve a better performance than CVP as it takes both noise caused interference and delivery imperfections into account, and it can reap the diversity gain as the SQR gets better with an increased SNR.



\vspace{12pt}


\begin{thebibliography}{00}
\bibitem{b1} G. Caire and S. Shamai, ``On the achievable throughput of a multiantenna Gaussian broadcast channel," \textit{IEEE Trans. Inf. Theory}, vol. 49, no. 7, pp. 1691-1706, July. 2003

\bibitem{b2} G. Caire and S. Shamai, ``On the achievable throughput of a multiantenna Gaussian broadcast channel," \textit{IEEE Trans. Inf. Theory}, vol. 49, no. 7, pp. 1691-1706, July. 2003

\bibitem{b3} M. Costa, ``Writing on dirty paper," \textit{IEEE Trans. Inf. Theory}, vol. 29, no. 5, pp. 439-441, May. 1983.

\bibitem{b4} A. Wiesel, Y. C. Eldar and S. Shamai, ``Zero-Forcing Precoding and Generalized Inverses," \textit{IEEE Trans. Signal Process.}, vol. 56, no. 9, pp. 4409-4418, Sept. 2008.

\bibitem{b5} T. M. Kim, F. Sun and A. J. Paulraj, ``Low-Complexity MMSE Precoding for Coordinated Multipoint With Per-Antenna Power Constraint," \textit{IEEE Signal Process. Lett.}, vol. 20, no. 4, pp. 395-398, April 2013.

\bibitem{b6}C. B. Peel, B. M. Hochwald and A. L. Swindlehurst, ``A vector-perturbation technique for near-capacity multiantenna multiuser communication-part I: channel inversion and regularization," \textit{IEEE Trans. Commun.}, vol. 53, no. 1, pp. 195-202, Jan. 2005.

\bibitem{b7}B. M. Hochwald, C. B. Peel and A. L. Swindlehurst, ``A vector-perturbation technique for near-capacity multiantenna multiuser communication-part II: perturbation," \textit{IEEE Trans. Commun.}, vol. 53, no. 3, pp. 537-544, March 2005.

\bibitem{b8} C. Windpassinger, R. F. H. Fischer, T. Vencel and J. B. Huber, ``Precoding in multiantenna and multiuser communications," \textit{IEEE Trans. Wireless Commun.}, vol. 3, no. 4, pp. 1305-1316, July 2004.

\bibitem{b9} M. Taherzadeh, A. Mobasher and A. K. Khandani, ``Communication Over MIMO Broadcast Channels Using Lattice-Basis Reduction," \textit{IEEE Trans. Inf. Theory}, vol. 53, no. 12, pp. 4567-4582, Dec. 2007.

\bibitem{b10} A. Ghasemmehdi and E. Agrell, ``Faster Recursions in Sphere Decoding," \textit{IEEE Trans. Inf. Theory}, vol. 57, no. 6, pp. 3530-3536, June 2011.

\bibitem{b11} D. A. Schmidt, M. Joham, and W. Utschick, ``Minimum mean square error vector precoding," \textit{Eur. Trans. Telecommun.}, vol. 19, no. 3, pp.219-231, Apr. 2008.

\bibitem{b12} P. Lu and H. Yang, ``Vector Perturbation Precoding for MIMO Broadcast Channel with Quantized Channel Feedback," in \textit{Proc. IEEE Global Telecommun. Conf.}, Nov./Dec. 2009, pp. 1-5.

\bibitem{b13}S. P. Herath, D. H. N. Nguyen and T. Le-Ngoc, ``Vector Perturbation Precoding Under Quantized CSI," \textit{IEEE Trans. Veh. Technol.}, vol. 65, no. 1, pp. 420-427, Jan. 2016.

\bibitem{b14} C. Masouros, M. Sellathurai and T. Ratnarajah, ``Limited feedback vector perturbation precoding by MinMax optimization," in \textit{Proc. IEEE Global Telecommun. Conf.}, Dec. 2014, pp. 3349-3353.
\bibitem{bb} J. Maurer, J. Jalden, D. Seethaler and G. Matz, ``Vector Perturbation Precoding Revisited," \textit{IEEE Trans. Signal Process.}, vol. 59, no. 1, pp. 315-328, Jan. 2011.



\bibitem{book}D. Tse and P. Viswanath, ``Fundamentals of Wireless Communication," Cambridge University Press, 2005.

\bibitem{wf}M. Joham, W. Utschick and J. A. Nossek, ``Linear transmit processing in MIMO communications systems," \textit{IEEE Trans. Signal Process.}, vol. 53, no. 8, pp. 2700-2712, Aug. 2005.

\bibitem{b15} T. L. Marzetta and B. M. Hochwald, ``Fast transfer of channel state information in wireless systems," \textit{IEEE Trans. Signal Process.}, vol. 54, no. 4, pp. 1268-1278, April 2006.

\bibitem{3gpp} 3GPP TS 36.211:``Evolved Universal Terrestrial Radio Access (E-UTRA); Physical channels and modulation"

\bibitem{b16} J. Salo, G. Del Galdo, J. Salmi, P. Kyösti, M. Milojevic, D. Laselva,and C. Schneider. (2005, Jan.) MATLAB implementation of the 3GPP Spatial Channel Model (3GPP TR 25.996) [Online]. Available:http://www.tkk.fi/Units/Radio/scm/ 

\bibitem{b17} I. E. Kaporin and I. N. Konshin, ``Parallel Solution of Symmetric Positive Definite Systems Based on Decomposition into Overlapping Blocks," \textit{Comput. Math. Math. Phys.}, vol. 41, pp. 481–493, Jan. 2001.

\end{thebibliography}
\end{document}